# Large Negative Flattened Dispersion over the S+C+L Bands Using Highly Birefringent Photonic Crystal Fiber

Md Borhan Mia[1*], Kanan Roy Chowdhury[2], Animesh Bala Ani[1], and Mohammad Faisal[1]
[1]Bangladesh University of Engineering and Technology, Dhaka, Bangladesh
[2]Chittagong University of Engineering and Technology, Chittagong, Bangladesh
*mborhanm1110@gmail.com

*Abstract*—Novel triangular lattice photonic crystal fiber (PCF) having vastly negative trodden chromatic dispersion over the S+C+L bands with a very high birefringence has been presented in this paper. To investigate different optical and electrical properties associated with the proposed fiber, finite element method (FEM) is deployed. The fiber presents a flattened negative dispersion of −698.5 ± 5 ps/(nm-km) over the wavelength of 1440 nm to 1600 nm. Besides, the proposed PCF displays a high birefringence of $1.886 \times 10^{-2}$ at 1550 nm wavelength. Furthermore, the nonlinearity, single modeness, effective area etc. of the proposed PCF are thoroughly discussed. The fiber would have important applications in broadband residual dispersion compensation as well as polarization maintaining applications

## I. INTRODUCTION

Recently high and ultra-high bit data are being transported by using dense wavelength division multiplexed (DWDM) and wavelength division multiplexed (WDM) systems [1]. In WDM system, bit rate of 40 Gb/s has been widely used [2]. Due to rapid grow of demand, 100 Gb/s [3] and 400 Gb/s [4] bit rate are employed to transport data for long haul communication. Standard single mode fibers (SMFs) are deployed in the transmission line which cause a huge stockpiled chromatic dispersion along the fiber optic line. Dispersion compensating fibers (DCFs) are used to recompense the accumulated dispersion of the SMFs. However, after compensation there exist some dispersion, known as residual dispersion due to slope mismatch arisen from the positive dispersion slope of SMFs. Therefore, residual dispersion compensating fibers (RDCFs) are indispensable to recompense the residual dispersion of the conventional DCFs. Besides, RDCFs exhibit very large negative and flattened dispersion, concurrently. There are many tunable properties of PCFs which are controllable and flexible. Therefore, PCFs have become very reasonable to design RDCFs.

Lately, PCFs with high negative flattened dispersion as well as large birefringence have been suggested by many authors. For instance, RDCF presented in [5] exhibits average negative dispersion of −457.4 ps/(nm-km) covering from 1360 nm to 1690 nm wavelength band. However, there is no information regarding birefringence and loss of the suggested PCF. In [6], Habib et al. proposed an octagonal shaped PCF. It shows a negative dispersion of −465.5 ps/(nm-km) and an out-and-out variation of 10.5 ps/(nm-km). Though, the design displays a high birefringence of $2.68 \times 10^{-2}$, the chromatic dispersion is small and the variation is high. Besides, the structure is hybrid which is difficult to fabricate. A PCF is suggested for residual dispersion purpose in [7] showing average negative dispersion of −138 ps/(nm-km) in the wavelength extending from 980 nm to 1580 nm with an out-and-out deviation of 12 ps/(nm-km). However, birefringence issue is ignored and flattened negative dispersion is very low. Ani et al. [8] proposed a RDCF exhibiting a flattened dispersion of −124 ps/(nm-km) covering the wavelength of 1350 to 1700 nm. Besides, fiber displays a dispersion deviation of ± 0.1252 ps/(nm-km). However, birefringence issue was disregarded. Li et al. [9] proposed a design, displaying a dispersion of −611.9 ps/(nm-km) in 1460 nm to 1625 nm wavelength bands. It also exhibits dispersion of −474 ps/(nm-km) over 1425 nm to 1675 nm wavelength. Hasan et al. [10] proposed photonic crystal fiber of spiral shape for broadband remaining dispersion reparation. The fiber shows flattened dispersion of −526.99 ps/(nm-km) over 1050 nm to 1700 nm with an absolute variation of ±3.7 ps/(nm-km) and birefringence of $2.26 \times 10^{-2}$. In [11], Hasan et al. proposed a photonic crystal fiber having high nonlinearity and high birefringent for dispersion purpose. Design fiber shows very high negative dispersion varying from −388.72 to −723.1 ps/(nm-km) in the wavelength array of 1460 to 1625 nm. However, transmission of data will be affected due to high nonlinearity of the suggested fiber.

In our article, we numerically investigated a triangular frame photonic crystal fiber for residual dispersion purpose. The fiber demonstrates a very high trodden dispersion of −698.5 ps/(nm-km) with a total variation of 5 ps/(nm-km) in S+C+L wavelength bands. To our knowledge, this is the utmost average negative dispersion than all recently printed articles. Moreover, the fiber exhibits a high birefringence of $1.885 \times 10^{-2}$ at the communication band, which will have important applications in sensor designs. Furthermore, the value of V-parameter recommends that fiber will operate only in fundamental mode.

## II. FIBER DESIGN

The geometric view of the designed PCF is demonstrated in Fig. 1. The fiber comprises of 7 rings of circular air holes, divided into two regions, i.e., core and cladding. The cladding comprises of 5 air hole rings and core has two air hole rings without any doping to keep the design simple. The outer cladding has a pitch value of $\Lambda_2 = 0.671\,\mu m$ and at the core region, the pitch value is $\Lambda_1 = 0.667\,\mu m$. The diameter of the air hole, both in core and cladding profile is $d_1 = 0.58\,\mu m$. The air filling fractions at core and cladding are 0.87 and 0.864, respectively. Three air holes are removed from the center which breaks the symmetry in the core. Asymmetric core is the prerequisite to obtain high birefringence. In the design, we used silica (SiO$_2$) as background material which is accessible in fiber industry.

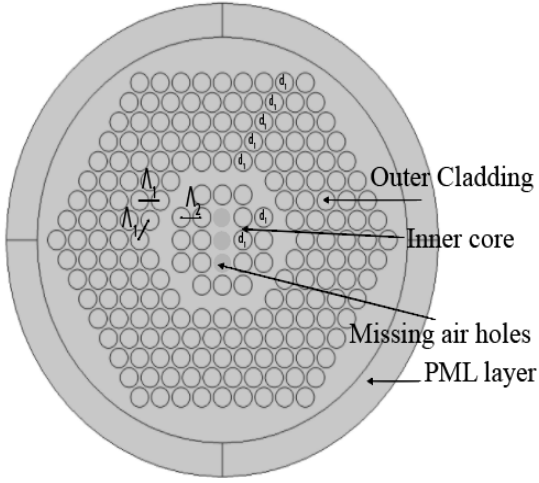

Fig. 1. Geometric view of the proposed PCF

### III. SIMULATION AND RESULT

Different belongings of the suggested fiber are tailored by using finite element method (FEM) and as a design simulator, COMSOL MULTIPHYSICS 5.0 is used. Perfectly matched layer (PML) is used to make simulation window compact and it produces no reflection. PML layer is placed nearby the outmost ring. Eigen value problems of Maxwell curl equation are solved to obtain the operative refractive index of the recommended fiber. Once $\eta_{eff}$ is obtained, other parameters; i.e., chromatic dispersion $D(\lambda)$, birefringence $B$, non-linear co-efficient $\gamma$, loss $L_c$, effective area $A_{eff}$ can be tailored from their respective equations [12]–[17]

$$D(\lambda) = -\frac{\lambda}{c}\frac{d^2 \text{Re}[\eta_{eff}]}{d\lambda^2} \quad (1)$$

$$B = |\eta_{eff}^x - \eta_{eff}^y| \quad (2)$$

$$A_{eff} = \frac{\left(\iint |E|^2 dxdy\right)^2}{\iint |E|^4 dxdy} \quad (3)$$

$$L_c = \frac{20\times 10^6}{\ln(10)} k_0 lm[\eta_{eff}] \quad (4)$$

$$V_{eff} = \frac{2\pi\Lambda}{\lambda}\sqrt{\eta_{eff}^2 - \eta_{FSM}^2} \quad (5)$$

Where, $Re[\eta_{eff}]$ and $Im[\eta_{eff}]$ are the real and imaginary fragment of the effective refractive index, respectively; $\eta_{eff}^x$ and $\eta_{eff}^y$ are the effective refractive indices alongside the x and y-polarization mode, respectively; c is the speed of light in void, $\lambda$ is the functional wavelength, $E$ is the electric field, $k_0 = \frac{2\pi}{\lambda}$, is the wave number in vacuum; $\eta_{eff}$ and $\eta_{FSM}$ are the refractive indices of the basic mode and basic space filling mode, respectively. The three termed Sellmeier principle is straightly included in simulation.

Fundamental mode profiles (x and y) of the suggested PCF is illustrated in Fig. 2 using parameters $d_1$ = 0.58 µm, $\Lambda_1$ = 0.667 µm and $\Lambda_2$ = 0.671 µm at wavelength of 1550 nm. Since higher refractive index is realized in core, optical fields are well restrained in core than cladding. From Fig. 2, it is enlightened that mode field spreads along the y-direction due to missing holes compared to x, which results in high birefringence.

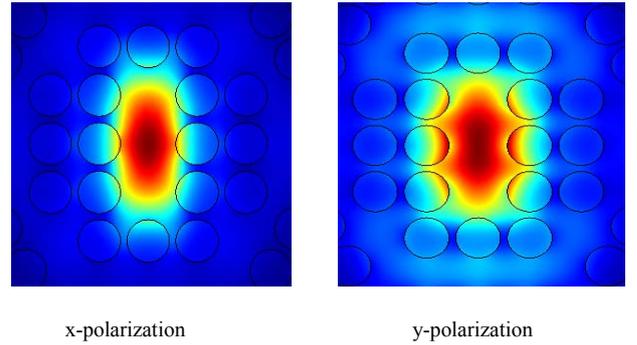

x-polarization          y-polarization

Fig. 2. Basic mode profile of the suggested PCF for x and y-polarization mode at 1550 nm.

In Fig. 3 chromatic dispersion of presented PCF is demonstrated as a function of wavelength for both x and y-polarization mode. Since, in highly birefringent PCF two polarization modes (x and y) exhibit unlike dispersion characteristics, we limit our discussion only for y-polarization mode which displays a large negative flattened dispersion compared to x-polarization mode. Proposed PCF shows a large dispersion of −698.5 ± 5 ps/(nm-km) in the wavelength band of 1440 nm to 1600 nm. The dispersion of the suggested PCF at 1550 nm is -700 ps/(nm-km) which is clearly depicted in Fig. 3.

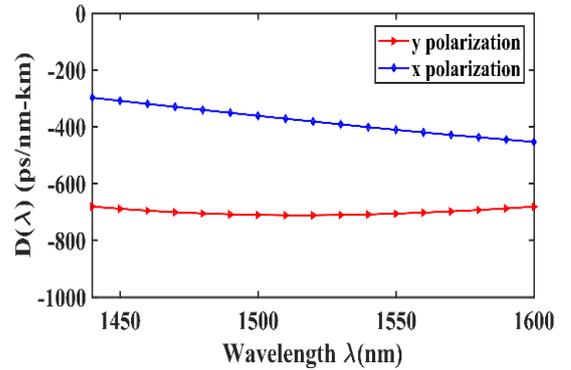

Fig. 3. Dispersion of proposed PCF as a function of wavelength alongside x and y polarization mode for the parameters $d_1$ = 0.58 µm, $\Lambda_1$ = 0.667 µm, $\Lambda_2$ = 0.671 µm.

To study, the effect of altering the diameter of the hole around the core region, we changed $d_1$ by taking the values of 0.57 µm, 0.58 µm and 0.59 µm and the corresponding chromatic dispersion is illustrated in Fig. 4. It is seen form the Fig. 4, when $d_1$=0.58 µm, a large negative flattened dispersion is obtained.

The effect of altering pitch, both inner core and outer cladding is demonstrated in Fig. 5 and Fig. 6, respectively. It is clearly seen form Fig. 5 and Fig. 6 that when $\Lambda_1$ = 0.667 µm and $\Lambda_2$ = 0.671 µm, we have attained a very large negative flattened dispersion from the bands of wavelength, 1440 nm to 1600 nm. Therefore, the optimum parameters of the proposed design are $d_1$ = 0.58 µm, $\Lambda_1$ = 0.667 µm and $\Lambda_2$ = 0.671 µm.

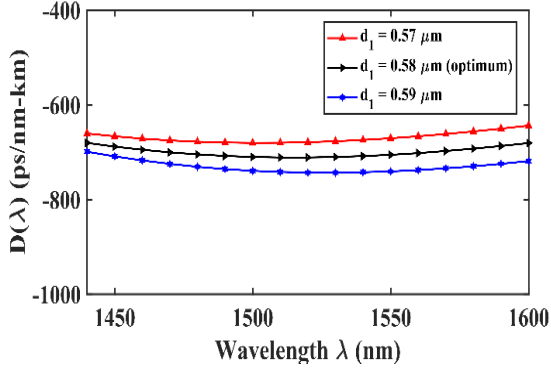

Fig. 4. Chromatic dispersion of the proposed PCF as function of wavelength altering $d_1$ keeping other parameters unaltered.

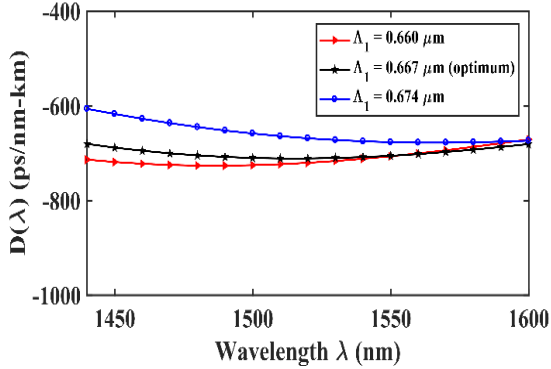

Fig. 5. Chromatic dispersion as a function of wavelength altering $\Lambda_1$, keeping other parameters unchanged.

The dispersion slope which is the variation of dispersion due to corresponding variation of wavelength, of the suggested PCF is demonstrated in Fig. 7 at optimum parameters. The dispersion slope is 0.35 ps/(nm$^2$-km) at communication band. Confinement loss of suggested PCF is also depicted in the same figure. In c band, confinement loss is 4×10$^{-5}$ dB/m, which is very low.

We carefully scrutinized birefringence, which is the variation between x and y-polarization mode, of the proposed fiber. To obtain birefringence, some air holes are intentionally removed from core to introduce asymmetry. Fig. 8 presents the birefringence of suggested PCF for finest factors $d_1$ = 0.58 µm, $\Lambda_1$ = 0.667 µm and $\Lambda_2$ = 0.671 µm. Form the Fig. 8, it is seen that, birefringence at 1550 nm wavelength is 1.885×10$^{-2}$, which is high enough compared to conventional polarization maintaining fibers.

The dependence of effective area on wavelength at finest parameters $d_1$, $\Lambda_1$ and $\Lambda_2$ is demonstrated in Fig. 9. The effective area of the suggested PCF is 4.2 µm$^2$ at 1550 nm wavelength and it decreases with the increment of the wavelength, due to guided mode of the fiber increases at longer wavelength. The nonlinear property of the proposed PCF is depicted in the same figure. The nonlinear co-efficient is 24.2 W$^{-1}$km$^{-1}$ at the exciting wavelength of 1550 nm. Since nonlinear co-efficient is low and it covers the wavelength of concern, the proposed fiber could be a proper choice for dispersion compensation.

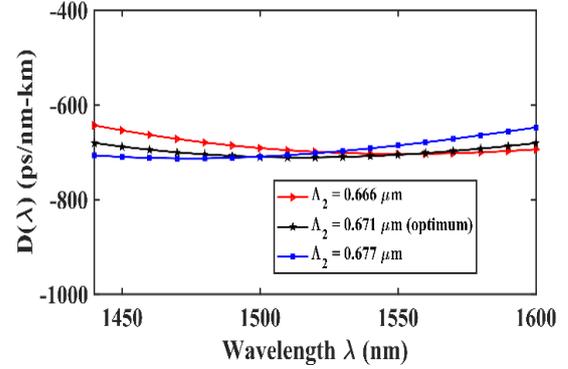

Fig. 6. Chromatic dispersion dependence on wavelength changing $\Lambda_2$, keeping other parameters constant.

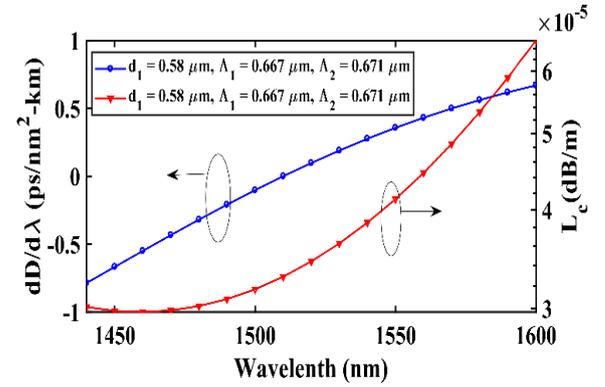

Fig. 7. Dispersion slope and loss as dependent on wavelength with parameters, $d_1$ = 0.58 µm, $\Lambda_1$ = 0.667 µm, $\Lambda_2$ = 0.671 µm.

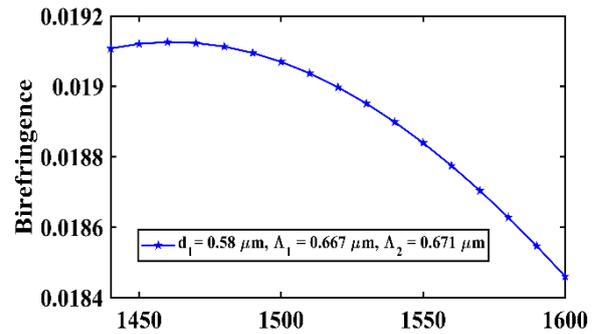

Fig. 8. Birefringence of the suggested fiber as a function of wavelength with parameters, $d_1$ = 0.58 µm, $\Lambda_1$ = 0.667 µm, $\Lambda_2$ = 0.671 µm.

The single modeness of the proposed PCF is carefully explored as discrepancy might happen because of different size of cores. The V- parameter assistances to verify whether the proposed fiber operates in single mode or not. The PCF will act as single mode fiber (SMF) if the V-parameter, $V_{eff} \leq \pi$. The V-parameter of the suggested PCF is showed in Fig. 10. From Fig. 10, it is realized that the V-parameter, $V_{eff} \leq \pi$ in wavelength band from 1440 nm to 1600 nm. Therefore, it guarantees that suggested PCF will operate in single mode only.

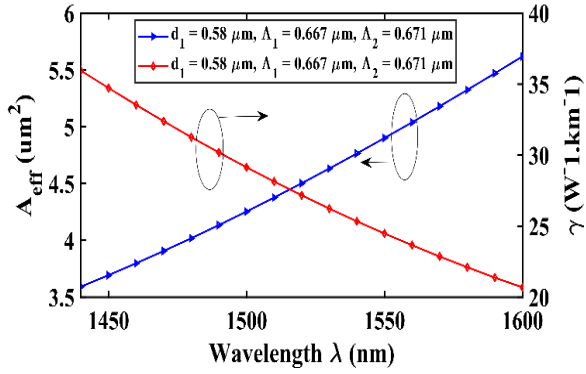

Fig. 9. Effective area and nonlinear co-efficient vs. wavelength with the finest parameters, $d_1 = 0.58$ μm, $Λ_1 = 0.667$ μm, $Λ_2 = 0.671$ μm.

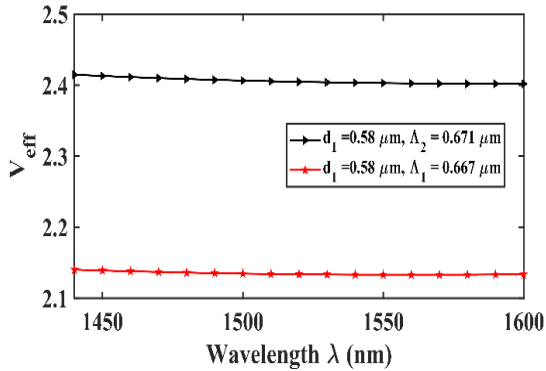

Fig. 10. V-parameter, $V_{eff}$ as a function of wavelength with optimum parameters, $d_1 = 0.58$ μm, $Λ_1 = 0.667$ μm, $Λ_2 = 0.671$ μm.

## IV. CONCLUSION

In this article, we demonstrated a novel PCF which shows a vast negative flattened dispersion over the wavelength S+C+L bands. The fiber exhibits large flattened dispersion of -698.5 ps/(nm-km) with a total variation of 5 ps/(nm-km). Therefore, fiber will be appropriate for dispersion compensation in optical transmission. Additionally, the proposed fiber exhibits large birefringence of $1.885×10^{-2}$ at 1550 nm wavelength, which has probable applications in polarization maintenance. Furthermore, the nonlinear co-efficient is 24.2 $W^{-1}km^{-1}$ at the wavelength of 1550 nm confirming that transmission of data won't be much affected by the nonlinearity of the suggested PCF. To our awareness, the suggested PCF exhibits the largest negative dispersion of all formerly published articles.